\documentclass[10pt, journal,twoside]{IEEEtran} 

\usepackage{graphicx}
\usepackage{subfigure}
\usepackage{amsmath}
\usepackage{amssymb}
\usepackage{cases}
\usepackage{times} 
\usepackage[font=footnotesize,labelfont=bf]{caption}
\usepackage{caption}
\usepackage{algorithm}
\usepackage{algorithmic}
\usepackage{color}
\usepackage{multirow}
\usepackage{algorithm, algorithmic}

\makeatletter
  \newcommand\figcaption{\def\@captype{figure}\caption}
  \newcommand\tabcaption{\def\@captype{table}\caption}
\makeatother
\usepackage{setspace}
\usepackage[utf8]{inputenc}
\usepackage[T1]{fontenc}
\usepackage{array} 
\newcolumntype{C}[1]{>{\centering}p{#1}}
\usepackage{multicol}  
\usepackage{multirow} 

\begin{document}
\title{Membership Information Leakage in Federated Contrastive Learning}
\author{Kongyang Chen, Wenfeng Wang, Zixin Wang, Wangjun Zhang, Zhipeng Li, Yao Huang
\IEEEcompsocitemizethanks{
\IEEEcompsocthanksitem{Kongyang Chen is with Institute of Artificial Intelligence, Guangzhou University, China. He is also with Pazhou Lab, Guangzhou, China. E-mail: kychen@gzhu.edu.cn.}
\IEEEcompsocthanksitem{Wenfeng Wang and Zixin Wang are with Institute of Artificial Intelligence, Guangzhou University, China.}
\IEEEcompsocthanksitem{Wangjun Zhang is with School of Computer Science and Cyber Engineering, Guangzhou University, China.}
\IEEEcompsocthanksitem{Zhipeng Li is with Jiangxi Qiushi Academy for Advanced Studies, Nanchang, China.}
\IEEEcompsocthanksitem{Yao Huang is with Guangzhou College of Commerce, Guangzhou, China. (Corresponding author: Yao Huang)}
}}

\IEEEtitleabstractindextext{
\begin{abstract}
Federated Contrastive Learning (FCL) represents a burgeoning approach for learning from decentralized unlabeled data while upholding data privacy. In FCL, participant clients collaborate in learning a global encoder using unlabeled data, which can serve as a versatile feature extractor for diverse downstream tasks. Nonetheless, FCL is susceptible to privacy risks, such as membership information leakage, stemming from its distributed nature, an aspect often overlooked in current solutions. This study delves into the feasibility of executing a membership inference attack on FCL and proposes a robust attack methodology. The attacker's objective is to determine if the data signifies training member data by accessing the model's inference output. Specifically, we concentrate on attackers situated within a client framework, lacking the capability to manipulate server-side aggregation methods or discern the training status of other clients. We introduce two membership inference attacks tailored for FCL: the \textit{passive membership inference attack} and the \textit{active membership inference attack}, contingent on the attacker's involvement in local model training. Experimental findings across diverse datasets validate the effectiveness of our attacks and underscore the inherent privacy risks associated with the federated contrastive learning paradigm.
\end{abstract}
}
\maketitle
\IEEEdisplaynontitleabstractindextext
\IEEEpeerreviewmaketitle

\section{Introduction}
In recent times, Federated Learning (FL) has emerged as a prominent field in the domain of decentralized data learning while upholding data privacy standards ~\cite{mcmahan2017communication}. In FL, individual client participants retrieve a global model from a centralized server and generate their own local models using their private data. These local models are then aggregated by the central server in an iterative manner to update the global model until convergence is achieved. By focusing on sharing model parameters instead of raw data, FL provides enhanced data privacy protection, making it a widely adopted approach in various sectors~\cite{yang2019federated, rieke2020future}. Nonetheless, FL operates under the assumption that each client possesses an ample amount of labeled data, which is often impractical. In reality, the availability of labeled data is significantly lower than that of unlabeled data, and annotating all unlabeled data is a time-consuming task, often rendering it infeasible.

Fortunately, the introduction of Federated Contrastive Learning (FCL)~\cite{van2020towards} has provided a solution to this challenge. In FCL, each decentralized client engages in collaborative learning to develop an encoder based on its local unlabeled data. This learned encoder can then serve as a feature extractor for constructing models for various downstream tasks using a limited amount of labeled data. As an innovative approach, FCL maximizes the utilization of distributed unlabeled data while safeguarding data privacy. Recent research efforts in the field of FCL have primarily concentrated on enhancing performance and scalability on non-IID data~\cite{zhang2020federated,zhuang2021collaborative,zhuang2022divergence}. Furthermore, several scholars have expanded the scope of FCL applications. For instance, Dong et al.~\cite{dong2021federated} and Wu et al.~\cite{wu2021federated} explored the extraction of representations from medical images under FCL settings, achieving notable results. Similarly, Saeed et al.~\cite{saeed2020federated} investigated the extraction of representations from unlabeled sensor data within IoT domains.

However, in FCL, little attention has been devoted to potential privacy risks, such as membership information leakage. Within the domain of artificial intelligence security, it is widely recognized that FL is susceptible to various security threats, including membership inference attacks, as extensively studied by numerous researchers~\cite{7958568, 8835245}. When considering membership inference attacks against FCL in comparison to FL, two key distinctions emerge. Firstly, FL's model training typically relies on labeled data, enabling attackers to utilize this label information for membership inference attacks. In contrast, FCL participants engage in unsupervised learning of an encoder without access to labeled local data, thereby making membership inference attacks notably more challenging. Secondly, the distributed nature and learning objectives of FCL diverge from those of FL. While FL aims to train a model for a specific task, FCL strives to develop a general encoder suitable for multiple downstream tasks. Given the potential variation in data distribution across these tasks, devising strategies for membership inference attacks based on data distribution becomes less viable.

This paper investigates the feasibility of executing a membership inference attack against FCL and introduces a viable strategy for launching such attacks. Given the decentralized nature of federated contrastive learning, a potential attacker may operate from a client-side perspective. Consequently, all actions of the attacker are confined to the local client, precluding the ability to manipulate the server-side aggregation process or gain insights into the training status of other clients. \textit{The attacker's objective is to determine whether the data represents training member information by accessing the model's inference output.} We propose two methods for conducting membership inference attacks: \textit{passive and active approaches}. Passive attackers abstain from participating in training or causing disruptions; instead, they solely acquire the model parameters post-training and then initiate membership inference attacks. In contrast, active attackers engage in gradient ascent on their local model, apply gradient ascent on the data to be inferred, transmit it to the server alongside other aggregated model parameters, and subsequently compute the loss for the data subjected to gradient ascent.

In particular, we delved into the \textit{passive membership inference attack} method at three distinct levels: the cosine similarity-based attack, the internal model-based attack, and the feature combination-based attack. Initially, membership inference is deduced through cosine similarity values. Subsequently, we extract the internal encoders of the contrastive learning model (typically a deep neural network like ResNet) and subject them to conventional membership inference attacks, unveiling the inherent internal privacy vulnerabilities of the contrastive learning model. Finally, we concatenate the highest confidence value in cosine similarity, data loss, and model encoder prediction into a three-dimensional array, followed by training a linear classifier for membership/non-membership prediction. In the \textit{active membership inference attack}, we implement gradient ascent on the overfitted model and observe that the data loss speed post-gradient ascent is slower for member data, thereby accomplishing member/non-member differentiation. These varied membership inference attack methods offer diverse insights into assessing the privacy risks associated with federated contrastive models. Lastly, we assess our attack performances across various datasets, including SVHN~\cite{netzer2011reading}, CIFAR-10~\cite{CIFAR}, and CIFAR-100~\cite{CIFAR}. The experimental results validate the efficacy of our attacks and underscore the privacy leakage risks inherent in the federated contrastive learning framework.

The primary contributions of our paper are outlined as follows:
\begin{enumerate}
\item We demonstrate the potential leakage of membership information in FCL. To the best of our knowledge, this is the first study to investigate the impacts of membership inference attacks on task-agnostic federated learning.
\item We introduce two membership inference attacks targeting FCL: the passive membership inference attack and the active membership inference attack, contingent upon whether each attacker has engaged with the local model training.
\item We assess the performance of our attacks using various datasets. Experimental results validate the effectiveness of our approaches.
\end{enumerate}

The remainder of this paper are structured as follows: Section~\ref{sec:related} reviews the related literature. Section~\ref{sec:framework} outlines the general framework and security threats of FCL. Section~\ref{sec:method} details our membership inference attack strategies against FCL. Section~\ref{sec:evaluation} presents the experimental outcomes of our methods. Finally, Section~\ref{sec:conclusion} provides the concluding remarks of this study.

\section{Related Work}
\label{sec:related}
\subsection{Contrastive Learning (CL)} 
In recent years, contrastive learning has garnered significant attention due to its remarkable performance, which often matches or even surpasses that of supervised learning, despite being trained on unlabeled data. While sometimes categorized under self-supervised learning methods, contrastive learning differs from traditional approaches by eschewing proxy tasks and instead directly training the encoder~\cite{doersch2015unsupervised, pathak2016context}. The selection of proxy tasks is crucial, and an inappropriate choice can lead to subpar model performance~\cite{gidaris2018unsupervised, zhang2016colorful}. Furthermore, traditional self-supervised learning methods lack a universally effective validation method for assessing model performance~\cite{sermanet2018time, noroozi2016unsupervised}. With the emergence of contrastive learning, these challenges have been significantly addressed.
Contrastive learning involves calculating cosine similarity for training; it generates two augmented images from the training data as positive samples, while other data in the same batch serve as negative samples. The primary objective is to bring the output of positive samples closer while distancing them from the negative sample outputs. For instance, SimCLR~\cite{pmlr-v119-chen20j} uses an encoder to produce feature vectors, which undergo projection through a multi-layer perceptron (MLP) to calculate contrastive loss. The model is then trained through backpropagation of this loss, and at the end of training, the MLP is discarded, leaving the trained encoder for downstream tasks such as detection and segmentation. This method yields high-performing models; however, it heavily relies on batch size, with a larger batch size leading to more negative samples and consequently better model performance.
An alternative approach involves utilizing a memory bank~\cite{wu2018unsupervised, misra2020self} to replace feature representations of negative samples without requiring an increase in the training batch size. Nonetheless, employing a memory bank entails significant computational expenses. To mitigate this, MoCo~\cite{He_2020_CVPR} introduced a momentum encoder that maintains a dictionary as a data sample queue, effectively increasing the batch size by using cached embeddings from historical steps. This reduces the need for large batch sizes in a single input and enhances computational efficiency. Similarly, BYOL~\cite{grill2020bootstrap} exclusively trains the encoder using positive samples.

\subsection{Federated Contrastive Learning (FCL)} 
Federated learning~\cite{pmlr-v54-mcmahan17a} is a distributed machine learning method that enables multiple devices to collaborate on model training using labeled data without the necessity of sharing raw data. In the context of unlabeled data, researchers have also delved into distributed contrastive learning methods and introduced privacy-preserving federated contrastive learning (FCL) strategies~\cite{Li_2021_CVPR}. These strategies maintain raw data locally while training a global encoder through the integration of local encoders. Recent studies primarily address the challenge of non-independent identically distributed (Non-IID) data in FCL.
For example, Zhang et al.~\cite{DBLP:journals/corr/abs-2010-08982} devise a dictionary module and an alignment module to enhance feature representation quality in Non-IID data settings. Similarly, Zhuang et al.~\cite{zhuang2021collaborative} introduce a divergence-aware module to mitigate weight divergence issues in Non-IID data scenarios. They leverage local knowledge to alleviate the Non-IID problem effectively. 
Huang~\cite{arxiv2023bafcl} centers on the analysis of backdoor attacks targeting FCL. This research stands as the inaugural investigation into the security vulnerabilities within the FCL framework, thus motivating our investigation into the potential leakage of membership information within the FCL domain.
Furthermore, contemporary solutions have explored the application of FCL across diverse domains~\cite{dong2021federated, wu2021federated, saeed2020federated}.

\subsection{Member Inference Attack (MIA)} 
Shokri et al.~\cite{7958568} initially introduce the concept of member inference attacks in federated learning, with the goal of ascertaining whether a given data point was part of a machine learning model's training set. Their method involves launching member inference attacks against deep learning models to infer the membership of training data through analysis of the model's prediction results. These attacks have the potential to disclose sensitive information, such as user privacy. Shokri et al.~\cite{8835245} further delineate passive and active member inference attacks.
More recently, Liu et al.~\cite{10.1145/3460120.3484749} propose EncoderMI, a member inference attack on contrastive learning. This approach achieves member inference attacks by training a binary classifier based on the discrepancy in cosine similarity between member data and non-member data in augmented data. 
Subsequent research has expanded the scope of member inference attacks, applying them to different types of machine learning models and scenarios, including model architecture privacy~\cite{scif2022privacy}, machine unlearning~\cite{arxiv2021unlearning, snn2023zhou, annal2023unlearning}, among others.

\subsection{Our Approach}
In contrast to existing solutions, our aim is to investigate membership inference attacks for unlabeled data using a distributed approach. To the best of our knowledge, this represents the first attempt to explore membership leakage issues in FCL.

\section{Membership Information Leakage in FCL}
\label{sec:framework}
In this section, we present the system architecture of FCL and briefly examine the issue of membership information leakage within FCL.

\subsection{Federated Contrastive Learning}
\begin{figure}[!t]
    \centering
    \includegraphics[width=0.7\linewidth]{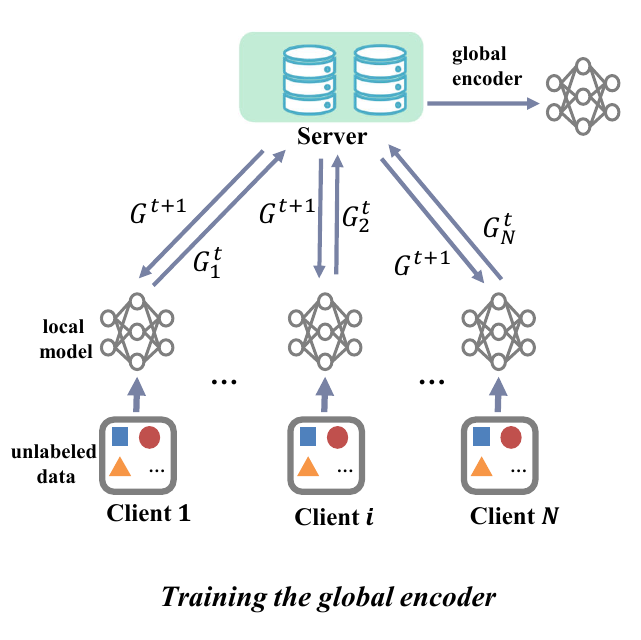}
    \caption{System architecture of FCL.}
    \label{fig:fcl-framework}
\end{figure}

FCL is a label-free distributed learning framework, as illustrated in Figure~\ref{fig:fcl-framework}. Conventional supervised learning necessitates manual annotation of numerous labels, incurring significant human resource costs. In contrast, unlabeled data, being easier to amass and obtain in larger quantities, offer an advantage.

\subsubsection{Server Parameter Aggregation} 
In FCL, each client initially receives the model structure and parameters from the server, proceeding to train locally with their respective unlabeled data $D_i$. With $N$ clients participating, they collaboratively train an encoder under the server's coordination. The objective of federated contrastive learning is to minimize the average loss across all client models. The loss function can be formulated as:
\begin{equation}
	\arg\min_{l \in \mathbb{R}^d} f(l) = \frac{1}{N} \sum_{i=1}^{N} f_i(l),
\end{equation}
\noindent where $f_i(\cdot)$ represents the loss function for the $i$-th client. Specifically, during the $t$-th round, the server transmits the current encoder $G_t$ to $n$ designated clients. Each designated client autonomously trains the server's encoder $G_t$  on its local unlabeled data $D_i$, resulting in the locally trained encoder $L^i_{t+1}$. Subsequently, the client sends the encoder update, $L^i_{t+1}-G_t$, back to the server. The server aggregates the received model parameters to derive a new server-side encoder $G_{t+1}$, following the formula:

\begin{equation}
	G_{t+1} = G_t + \frac{\eta}{n} \sum_{i=1}^{n} (L_{t+1}^i - G_t)
\end{equation}

The training process continues iteratively until the server's encoder converges. \textit{FCL differs significantly from FL in several key aspects}. Firstly, FL focuses on modeling a specific task, whereas FCL aims to develop an encoder suitable for multiple downstream tasks. Secondly, while FL employs a supervised local training algorithm, FCL utilizes an unsupervised local training approach.

\subsubsection{Local Contrastive Learning}
Contrastive learning endeavors to maximize the similarity between augmented data of the same instance and minimize the similarity between outputs of different instances. In most existing contrastive learning frameworks, InfoNCE~\cite{van2018representation} serves as the primary training loss function. In our experiments, we adopt MoCo~\cite{He_2020_CVPR} for unsupervised learning, which seeks to learn discriminating features in the representation space for positive and negative examples. The core concept of MoCo involves incorporating a momentum encoder and a sizeable memory queue to construct the contrastive loss function.

The MoCo architecture comprises two encoders: a query encoder and a key encoder with momentum. The query encoder extracts feature representations $\hbar_q$ from the original data, while the key encoder, with a momentum update strategy, extracts feature representations $\hbar_k$. While the query encoder's parameters are updated at each training step, the key encoder's parameters evolve based on those of the query encoder with a specific momentum coefficient.
To create positive and negative pairs, MoCo leverages data augmentation techniques to generate two distinct views (i.e., two samples). Matching views from the same sample constitute a positive pair, whereas views from dissimilar samples form a negative pair. The memory queue retains previous keys, enabling consideration of additional negative samples when computing the contrastive loss. Each queue element represents a feature vector derived from encoding the original data.
For the positive pair (i, j), the loss function of MoCo is defined as:

\begin{equation}
	L(i, j) = -\log \left( \frac{\exp \left( f(q_i) \cdot f(k_j) / \tau \right)}{\sum_{n=1}^N \exp \left( f(q_i) \cdot f(k_n) / \tau \right)} \right).
\end{equation}
Here, $f(q_i$) and $f(k_j)$ represent the feature representations of the query and key, respectively. $N$ denotes the number of negative samples, and $\tau$ serves as the temperature parameter, smoothing the loss function. This function aims to maximize similarity between positive samples while minimizing similarity between negative samples. In comparison with SimCLR~\cite{pmlr-v119-chen20j}, MoCo incorporates more negative samples through the momentum encoder and memory queue into the loss function, demonstrating strong performance in unsupervised learning tasks by effectively extracting meaningful feature representations.

\subsection{Membership Information Leakage in FCL}
Given its distributed nature, FCL may face security concerns akin to those in FL, including membership information leakage. Membership inference attacks aim to discern whether a specific data sample was involved in the model's training process. Attackers scrutinize the global model to identify training data samples, potentially leading to data leaks and privacy breaches, particularly when the data entails sensitive information. 

\textit{However, membership inference attacks against FCL differ notably from those targeting FL}. Firstly, FL's model training typically relies on labeled data, enabling attackers to exploit this label information for membership inference attacks. In contrast, FCL participants learn an encoder in an unsupervised manner, devoid of labeled local data, rendering membership inference attacks more challenging. Secondly, the distributed nature and learning objectives of FCL diverge from those of FL. While FL aims to train a model for a specific task, FCL seeks to develop a general encoder applicable to multiple downstream tasks. As data distribution may vary across these tasks, strategies for membership inference attacks based on data distribution become less feasible.

\section{Membership Inference Attacks Against FCL}
\label{sec:method}
\subsection{The Attacker's Capabilities}
Due to the decentralized nature of federated contrastive learning, a potential attacker could originate from a client. \textit{The objective of the attacker is to ascertain whether the data represents training member data by accessing the model inference output.} We posit that the attacker is a member of the clients. They possess the capability to adjust their local model parameters and training data, and also hold some data conforming to the same distribution as the overall training data. Consequently, all of the attacker's actions are confined to the local client, and they lack the capacity to manipulate the server-side aggregation method or comprehend the training status of other clients.

Assuming the attacker originates from the client, they have two methods for conducting membership inference attacks, which we categorize as \textit{passive and active membership inference attacks}. Passive attackers do not engage in training or disrupt it; instead, they solely acquire the model parameters post-training and subsequently initiate membership inference attacks. Conversely, active attackers conduct gradient ascent on their local model, execute gradient ascent on the data to be inferred, transmit it to the server along with other aggregated model parameters, and subsequently compute the loss for the data subjected to gradient ascent.

\begin{figure}[!t]
    \centering
    \includegraphics[width=0.7\linewidth]{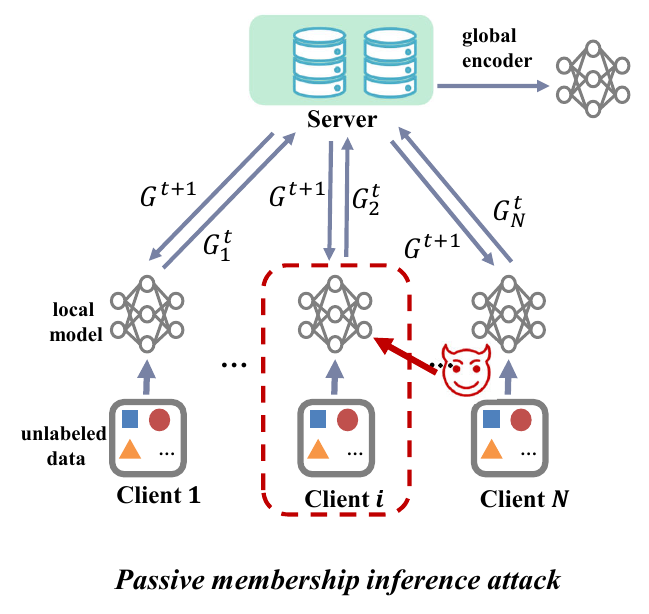}
    \caption{Passive membership inference attack on FCL: The attacker refrains from disrupting the FCL training process and solely acquires the aggregated model parameters for inference.}
    \label{fig:passive membership attack}
\end{figure}

\subsection{Passive Membership Inference Attack}
The system framework for the passive attack is illustrated in Figure~\ref{fig:passive membership attack}. Our approach is a modification of supervised membership inference attacks and contrastive learning membership inference attacks tailored for federated learning. The passive membership inference attack methodology can be categorized into three levels: \textit{the cosine similarity-based attack}, \textit{the internal model-based attack},  and \textit{the feature combination-based attack}.

\begin{figure*}[!t]
	\centering
	\includegraphics[width=0.8\textwidth]{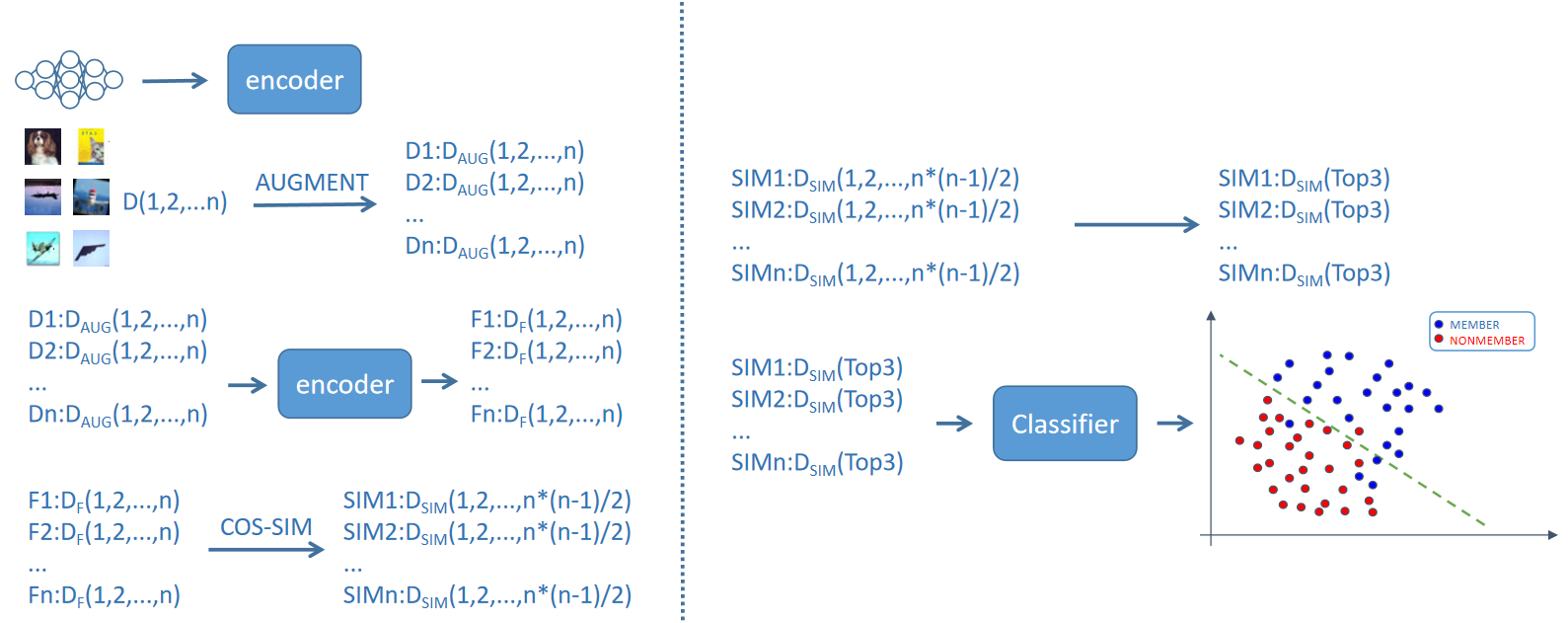}
	\caption{To generate augmented data, we start with the training dataset $D$. We randomly augment each data in $D$ to produce $n$ new augmented data. These augmented data are then fed into the model's encoder to obtain feature vectors $F$. Next, we calculate the cosine similarity $SIM$ among the augmented data from the same source. Finally, we use the top three values of the highest cosine similarity (Top3) for each data as the 3D features of the input data for training the binary classifier.}
	\label{fig:passive mia}
\end{figure*}

\subsubsection{Passive membership inference attack with cosine similarity}
The initial type entails extracting the encoder from the trained model, generating $M$ augmented data for each inference data, and feeding these augmented data into the encoder to acquire feature vectors. Subsequently, the cosine similarity for each feature vector is computed, and the average similarity $(S_x)$ of the sample with all samples in the dataset $D$ is calculated using the following formula:
\begin{equation}
	M(x, \tilde{h}) = \{S(\tilde{h}(x_i), \tilde{h}(x_j)) \mid i \in [1, n], j \in [1, n], j > i\}
\end{equation}
In the equation, $M(x, \tilde{h})$ denotes a set where $x$ represents the input sample and $\tilde{h}$  signifies the pre-trained encoder. This set captures the similarity scores of every possible pair of samples ($x_i$, $x_j$) within the sample dataset, with $i$ and  $j$ denoting the respective indices of the samples, and $n$ representing the total sample count in the dataset. The condition $j>i$ is imposed to prevent the inclusion of duplicate sample pairs.

Similar to supervised membership inference attacks, a binary classifier is trained and utilized for predicting outcomes on the data, as illustrated in Figure~\ref{fig:passive mia}. By computing similarity scores, we can assess the encoder's performance and glean insights into sample relationships that are beneficial for subsequent tasks. The formula can be expressed as follows:
\begin{equation}
	E = \{(M(x, \tilde{h}), 1) \mid x \in D_{sm}\} \cup \{(M(x, \tilde{h}), 0) \mid x \in D_{snm}\}
\end{equation}

The formula introduces the set $E$, which contains paired elements $((M(x, \tilde{h}), y)$. Here, $y$y denotes a binary label, where $y = 1$  if $x$ belongs to the trained sample set $D_{sm}$; and $y = 0$ if $x$ belongs to the untrained sample set $D_{snm}$. The subset $(M(x, \tilde{h})$ comprises similarity scores between all possible sample pairs mapped by the pre-trained encoder $\tilde{h}$  from the input sample  $x$.

\subsubsection{Passive membership inference attack with internal models}
The second approach involves using the encoder to conduct a generic membership inference attack, originally designed as a supervised membership inference attack. Here, the encoder predicts the label without requiring real labels. It only needs to be informed that there are $N$ categories, and then make a prediction. We then select the top $K$ data with the highest predicted probability as the input data for training the binary classifier. In this case, suppose the encoder is $G$, and the input data sample is $x$, the formula is as follows:
\begin{equation}
	f = \operatorname{ \textit{Train}}  (\operatorname{ \textit{Top K}}(\operatorname{ \textit{Predict}} (G(x), N), K))
\end{equation}

The predicted probabilities will exhibit a certain level of confidence, with the model showing very high probabilities for the first few predicted outcomes. Consequently, this method is capable of detecting model overfitting at an early stage. When the accuracy of other attacks is not optimal, this method can preemptively identify the risk of model overfitting. Furthermore, we can delve deeper by extracting the output of each layer in the encoder for membership inference attacks, as certain layers may achieve higher inference accuracy.

\subsubsection{Passive membership inference attack with feature combinations}
The third method involves constructing three-dimensional data by combining the cosine similarity of the data, the loss, and the highest prediction output probability. This three-dimensional data is then input into the classifier for training member and non-member data through binary classification. The structure of the classifier training data is as follows:
\begin{equation}
	(x, y, z) = (\operatorname{ \textit{CosineSimilarity}} (a, b), \operatorname{ \textit{Loss}} (c, d), \operatorname{ \textit{MaxProb}} (P))
\end{equation}

This approach can enhance the accuracy of distinguishing between member and non-member data by comparing three distinct features, thereby significantly improving overall accuracy.

\subsection{Active Membership Inference Attack}
The active membership inference attack is categorized into two types: one is conducted after the completion of federated contrastive learning training, and the other is performed during the federated contrastive learning process.

The first approach involves acquiring the model parameters after the conclusion of the FCL. Subsequently, this model is utilized to conduct gradient ascent on the data to be inferred, while observing the increase in data loss. The formula for gradient ascent is given below:
\begin{equation}
	x_{t+1} = x_t + \alpha \nabla J(x_t)
\end{equation}

\begin{algorithm} [!htb]
	\caption{Active Membership Inference Attack on Federated MoCo with Gradient Ascent}
	\begin{algorithmic}[1]
		\STATE \textbf{Input:} Target federated model $M$, member dataset $D_{member}$, inference dataset $D_{inference}$, threshold $T$, gradient ascent steps $G$, learning rate $\eta$
		
		\FOR {data batch $x_q$ in $D_{member} \cup D_{inference}$}
		\STATE Sample two random positive views $x_q^1$ and $x_q^2$ from $x_q$
		
		\STATE \textbf{Gradient ascent:}
		\FOR {$i=1$ to $G$}
		\STATE Compute query and key vectors for both views:
		\STATE $q^1 = M(x_q^1)$
		\STATE $k^1 = M(x_q^2)$
		\STATE $q^2 = M(x_q^2)$
		\STATE $k^2 = M(x_q^1)$
		
		\STATE Compute contrastive losses for both pairs:
		\STATE $L_1 = -\log(\frac{\exp(q^1 * k^1)}{\sum_{i=1}^K \exp(q^1 * Q_i)})$
		\STATE $L_2 = -\log(\frac{\exp(q^2 * k^2)}{\sum_{i=1}^K \exp(q^2 * Q_i)})$
		\STATE $L = L_1 + L_2$
		
		\STATE Perform gradient ascent with SGD and learning rate $\eta$ using $-L$
		\ENDFOR
		
		\STATE \textbf{Aggregate:}
		\STATE Send updated model parameters to the server
		\STATE Server aggregates the updated models
		\STATE Receive aggregated model from the server
		
		\STATE \textbf{Compute loss difference:}
		\STATE Compute MoCo contrastive loss for the aggregated model $L_{agg}$
		\STATE Compute loss difference $\Delta L = |L - L_{agg}|$
		
		\STATE \textbf{Membership inference:}
		\IF {$\Delta L < T$}
		\STATE Classify $x_q$ as a member
		\ELSE
		\STATE Classify $x_q$ as a non-member
		\ENDIF
		\ENDFOR
	\end{algorithmic}
\end{algorithm}

In this formula, $x_t$ represents the position at the $t$-th iteration; $\alpha$ denotes the learning rate, controlling the step size of each iteration update; $\nabla J(x_t)$ signifies the gradient of the objective function $J$ with respect to $x_t$; and $x_{t+1}$ denotes the new position after the $t+1$-th iteration. The gradient ascent operation leverages the discrepancy between the model's member and non-member data, where the loss of trained data increases more gradually compared to untrained data. This divergence can be utilized to infer whether a data point is a member or non-member data. The data undergoing gradient ascent in this context is the enhanced data intended for inference. Typically, the enhancement is random. If the enhanced data used differs significantly from that during training, it may impact the final result accuracy. However, in cases of severe overfitting, satisfactory accuracy can still be attained even with dissimilar enhanced data.

\begin{figure}[!t]
    \centering
    \includegraphics[width=0.7\linewidth]{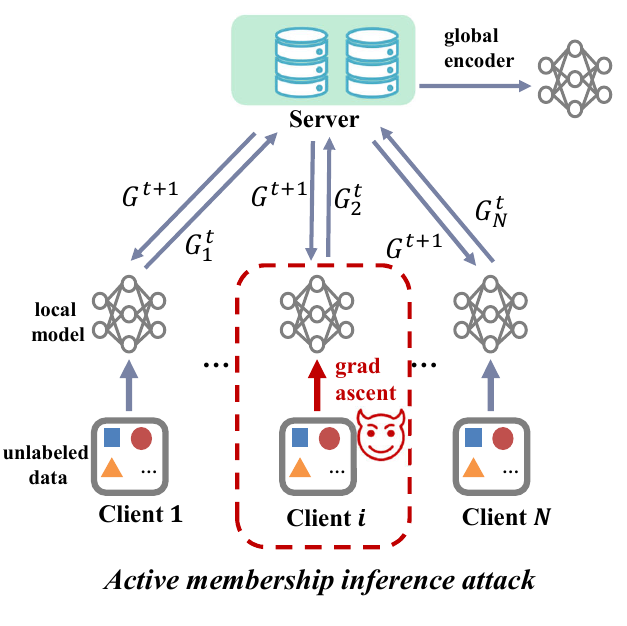}
    \caption{Active membership inference attack on FCL: The attacker uploads modified model parameters to expedite the inference process.}
    \label{fig:active_membership_attack}
\end{figure}

The second method assumes that the attacker is one of the client's members. If the attacker aims to determine whether a set of data is member data, they can adjust their own model, utilize it to conduct gradient ascent on the data for inference, and then upload the model to the server. Upon the server aggregating and returning the model parameters, the loss or cosine similarity of the data for inference is computed to observe its alterations. This process continues, recording the loss or cosine similarity of the model with the inferred data with each issuance. A fundamental assumption here is that the attacker must possess a small amount of member data for comparison. If the loss obtained post-gradient ascent is substantially lower when contrasted with the member data, it indicates that the inferred data is member data. Conversely, a significant disparity suggests that the inferred data is non-member data. Figure~\ref{fig:active_membership_attack} illustrates the flowchart of the active membership inference attack.

\section{Experiments}
\label{sec:evaluation}
\subsection{Experimental Settings}
\begin{table*}[!htb]
	\centering
	\caption{Datasets description.}
	\begin{tabular}{ccccc}
		\hline
		Datasets & Shape       & Classes & Number of Training data & Number of  of testing data \\ 
		\hline
		SVHN      & 32 $\times$ 32$\times$ 3     & 10      & 73,257    & 26,032    \\ 
		CIFAR-10  & 32 $\times$ 32$\times$ 3     & 10      & 50,000    & 10,000    \\
		CIFAR-100 & 32 $\times$ 32$\times$ 3     & 100     & 50,000    & 10,000    \\
		\hline
	\end{tabular}
	\label{tab:datasets}
\end{table*}

\subsubsection{Datasets} 
Our experiments were conducted on the SVHN~\cite{netzer2011reading}, CIFAR-10~\cite{CIFAR}, and CIFAR-100 datasets~\cite{CIFAR}, with the distribution presented in Table \ref{tab:datasets}. 

\begin{enumerate}
\item [a)]  The Street View House Numbers (SVHN) Dataset~\cite{netzer2011reading} comprises 73,257 samples for training and 26,032 samples for testing. Each sample is a $32 \times 32 $ RGB image depicting a digit number, coming from house numbers in Google Street View images.
\item [b)]  The CIFAR dataset~\cite{CIFAR}, sourced from real-world images like airplanes, birds, and cats, includes 60,000 RGB images sized at $32 \times 32$. Within this dataset, 5,000 images are allocated for training and 10,000 for testing purposes. CIFAR exists in two variations: CIFAR-10 and CIFAR-100. CIFAR-10 encompasses 10 classes with 6,000 samples per class, whereas CIFAR-100 consists of 100 classes, each with 600 images.
\end{enumerate}

\subsubsection{Experimental Details} 
The mentioned attacks were implemented using PyTorch in Python 3.7, leveraging 4 NVIDIA V100 GPUs. While each experiment was successfully carried out across all datasets, some attack diagrams lacked significant variation, thus CIFAR-10 diagrams were used to represent them. 
All experiments were executed under Non-IID distributions, as this aligns more with real-world federated learning scenarios, despite IID distribution being a more effective defense against overfitting.

In passive membership inference attacks, model parameters were logged every hundred aggregation rounds for subsequent inference analysis. To expedite overfitting, the initially provided server model parameters underwent 500 rounds of pre-training. Standardizing variables, 10,000 training and test data points were combined for all datasets, utilizing both member and non-member data during membership inference operations. For active membership inference attacks, losses of member and non-member data were tracked at each aggregation round, with model parameters saved every hundred rounds for static active membership inference analysis. Local training rounds were set at 10. Although the learning rate parameters for SGD during gradient ascent and training were both 0.1, adjusting the learning rate during gradient ascent appropriately helped mitigate large loss fluctuations, resulting in a smoother curve progression.

To assess the encoder's performance, typical methods include linear evaluation and weighted KNN evaluation. Linear evaluation gauges the feature representation extracted by the encoder through training a linear model. Meanwhile, weighted KNN evaluation classifies by comparing the cosine similarity of feature representations and utilizing a weighted voting k-nearest neighbor approach. Throughout training, weighted KNN evaluation was employed for monitoring, while linear evaluation was utilized for final performance testing.

\subsubsection{Evaluation Metrics} 
Passive membership inference attacks predominantly evaluate the classifier; hence, our evaluation metrics encompass accuracy, precision, and recall. \textit{Accuracy} denotes the ratio of correctly classified samples to the total sample count. \textit{Precision} signifies the proportion of samples predicted as the positive class by the classifier that actually belong to the positive class. It focuses on the accuracy of samples labeled as positive by the classifier. \textit{Recall} represents the proportion of samples correctly predicted as the positive class by the classifier among those that truly belong to the positive class. It pertains to the classifier's ability to identify positive samples.

For active membership inference attacks, the approach involves observing the rate of increase in member and non-member data loss and utilizes a threshold for inference. An appropriate loss threshold is applied to infer member and non-member data, where loss refers to the data loss after the increase minus the data loss before the increase. During the training process of the active membership inference attack, inference is made by comparing with the loss of member data. Specifically, a batch of data to be inferred undergoes gradient ascent, and the resulting loss is subtracted from the member data loss, then the difference is observed. A small difference indicates member data, whereas a large difference indicates non-member data. A threshold can also be set for the difference; if exceeded, it signifies non-member data, otherwise, it signifies member data. This method enables dynamic threshold adjustments based on specific scenarios and data characteristics.

It is crucial to note that while the aforementioned experimental setup and evaluations are structured, it is imperative to consider the ethical implications and privacy concerns associated with membership inference attacks when conducting such experiments. These methods should be leveraged to assess system robustness and security rather than exploiting potential vulnerabilities for unethical purposes.

\subsection{Experimental Results}

\begin{figure*}[!t]
\centering
\subfigure[CIFAR-10: loss function.]{
\includegraphics[width=0.3\linewidth]{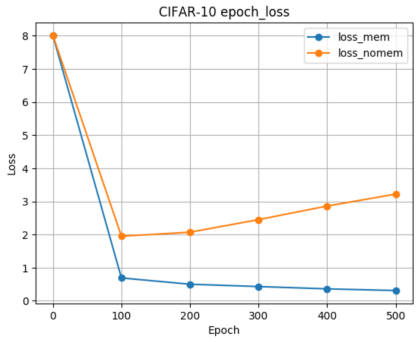}
\label{fig:passive membership1}
}
\subfigure[CIFAR-100: loss function.]{
\includegraphics[width=0.3\linewidth]{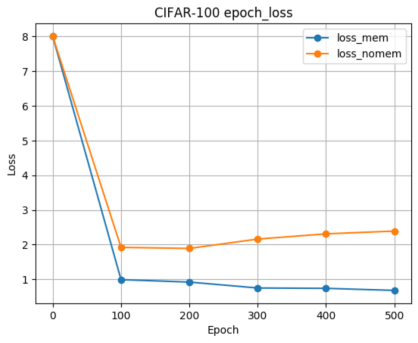}
\label{fig:active membership2}
}
\subfigure[SVHN: loss function.]{
\includegraphics[width=0.31\linewidth]{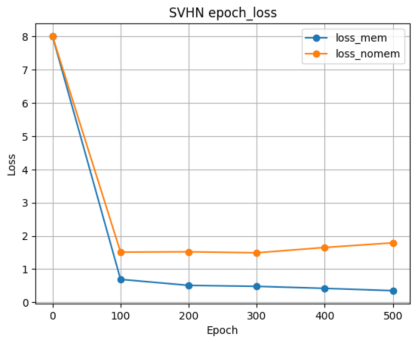}
\label{fig:active membership3}
}
\subfigure[CIFAR-10: cosine similarity.]{
\includegraphics[width=0.3\linewidth]{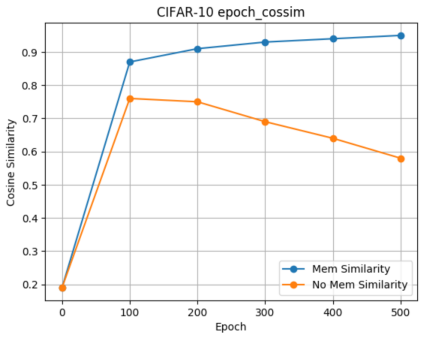}
\label{fig:passive membership4}
}
\subfigure[CIFAR-100: cosine similarity.]{
\includegraphics[width=0.3\linewidth]{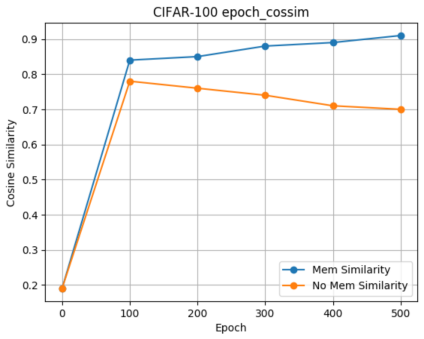}
\label{fig:active membership5}
}
\subfigure[SVHN: cosine similarity.]{
\includegraphics[width=0.3\linewidth]{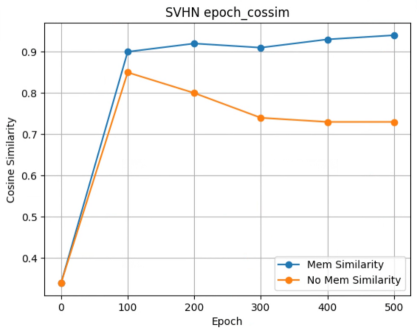}
\label{fig:active membership6}
}
\caption{The overfitting characteristics across different datasets.}
\label{fig:datatsets_data}
\end{figure*}

\subsubsection{Passive membership inference attack with cosine similarity}\label{passive-mis-cos}
During the passive membership inference attack, we initially utilized a method based on cosine similarity for the experiment. This method is primarily employed to identify the overfitting characteristics of FCL. The experimental findings substantiate that FCL does indeed harbor privacy risks stemming from overfitting. While the federated learning framework exhibits stronger resistance to overfitting compared to one single client, there remains a potential risk of overfitting. Table~\ref{tab:acc} showcases the performance data of membership inference attacks based on cosine similarity conducted on the model derived after multiple rounds of federated aggregation.

\begin{table}[!t]
	\centering
	\caption{Datasets description of membership inference attacks.}
	\begin{tabular}{cccc}
		\hline
		Datasets  & Accuracy & Precision & Recall \\ 
		\hline
		CIFAR-10  & 0.914    & 0.897     & 0.943  \\
		CIFAR-100 & 0.932    & 0.911     & 0.964  \\
		SVHN      & 0.905    & 0.901     & 0.924  \\
		\hline
	\end{tabular}
	\label{tab:acc}
\end{table}

Throughout model training, we gathered the model parameters after every hundred rounds of aggregation and utilized the model to compute the cosine similarity and loss for member and non-member data, as depicted in Figure~\ref{fig:datatsets_data}. 

It is evident that the model's generalization capability diminishes post overfitting. Despite the contrast learning model inherently possessing robust overfitting resistance, the model's generalization gradually deteriorates with an increase in the number of training epochs. Once the model becomes overfitted, the loss of non-member data progressively increases, indicating a decline in the model's generalization capacity. Similarly, as the model's overfitting intensifies, the cosine similarity of non-member data gradually decreases.

These results offer crucial insights into the privacy risks linked with federated contrastive learning due to overfitting. They underscore the necessity for further exploration of strategies to mitigate these risks while upholding the efficacy of the learning process.

\subsubsection{Passive membership inference attack with internal models}
\begin{figure*}[!t]
\centering
\subfigure[feat1.]{
\includegraphics[width=0.3\linewidth]{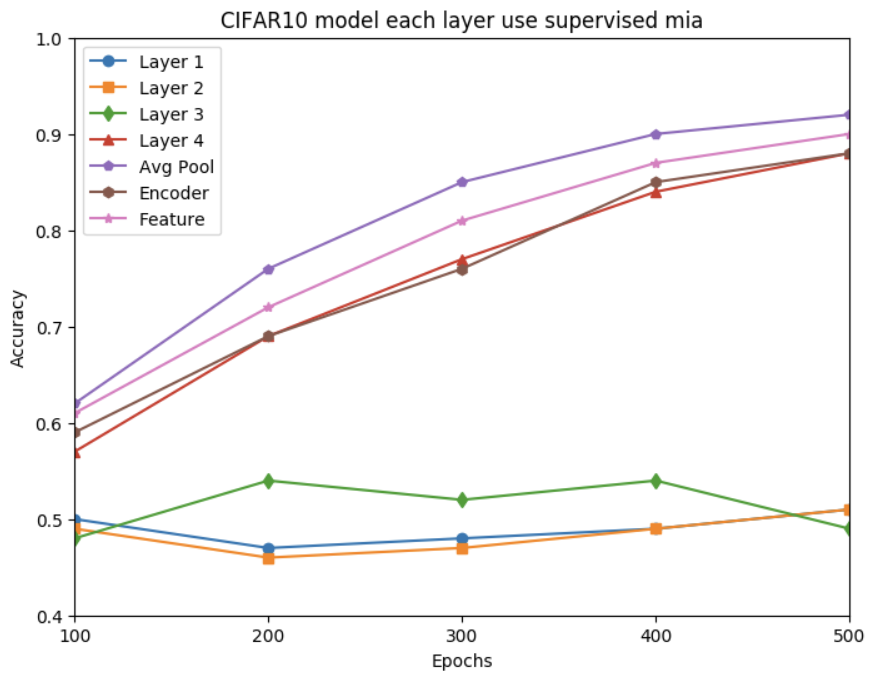}
\label{feat1}
}
\subfigure[feat2.]{
\includegraphics[width=0.3\linewidth]{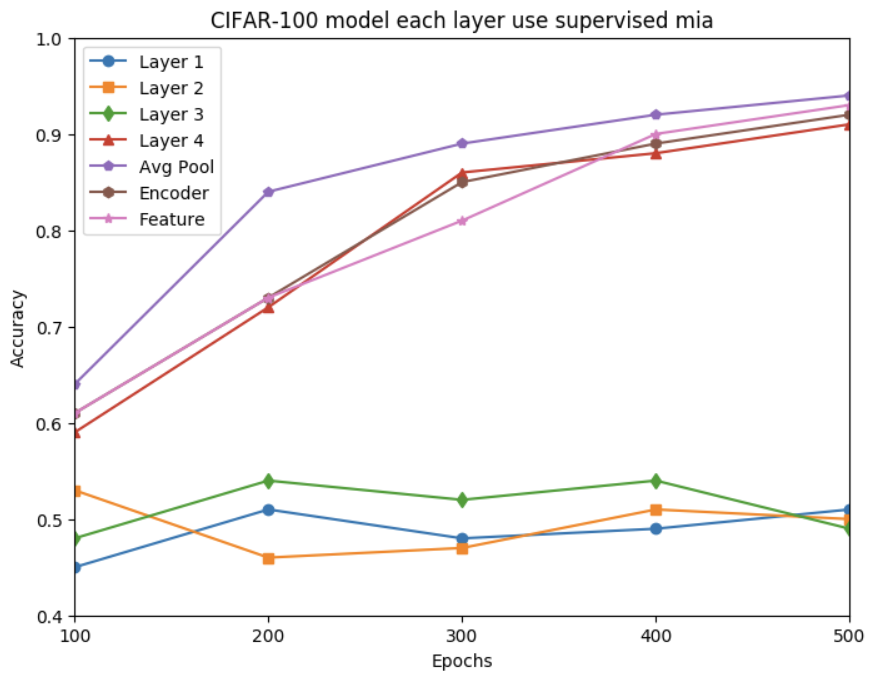}
\label{feat2}
}
\subfigure[feat3.]{
\includegraphics[width=0.3\linewidth]{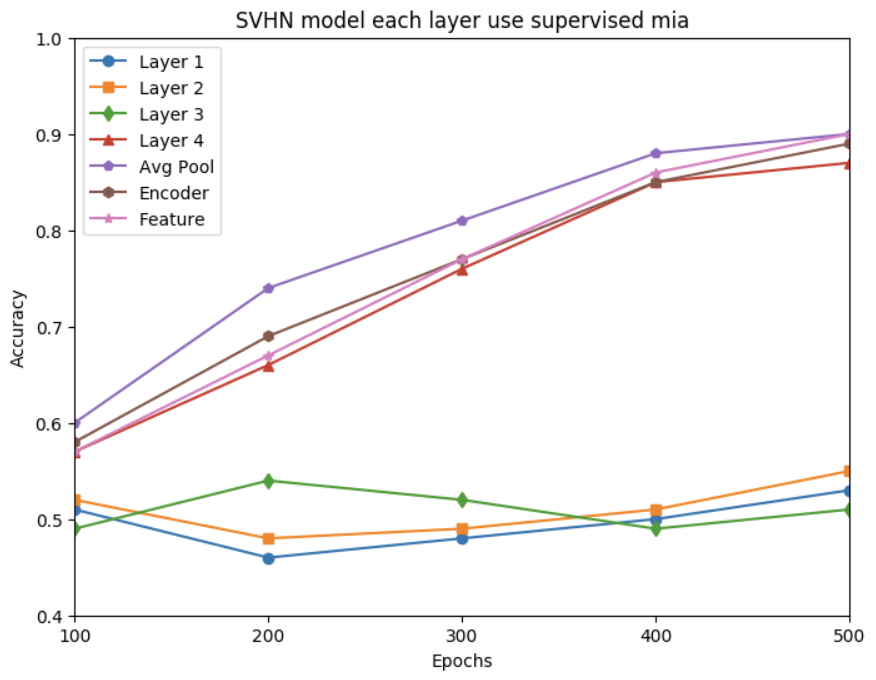}
\label{feat3}
}
\caption{The evaluation involves comparing each layer within the model's encoder with the membership inference attack utilizing cosine similarity (as discussed in Section~\ref{passive-mis-cos}). The accuracy of the membership inference attack on the model is logged every hundred epochs during model training. In this context, 'layer' and 'avgpool' pertain to the outputs of internal encoder layers, 'encoder' represents the aggregate output, and 'feature' denotes the accuracy associated with the initial membership inference attack type discussed earlier. 'Accuracy' signifies the classifier's precision in categorizing the test dataset.}
\label{fig:Contrast}
\end{figure*}

In the subsequent analysis, our focus shifts towards a generic membership inference attack predicated on the model's internal workings. Our experimental findings reveal that an overfitted contrast learning model leads to heightened overfitting within the extracted encoder. Given that this encoder typically constitutes a deep neural network model such as \textit{ResNet} or \textit{VGG}, we can directly apply traditional membership inference techniques to it. This process involves utilizing the model to predict class probabilities and selecting the highest probabilities as input data for training the classifier. By providing the model with the number of classes without specific labels, it generates prediction probabilities with a notable degree of confidence, reflected in significantly elevated values for the top predictions. Thus, we posit that this approach offers a viable avenue for further exploration of the model's privacy vulnerabilities.

\begin{figure*}[!t]
\centering
\subfigure[encoder.]{
\includegraphics[width=0.3\linewidth]{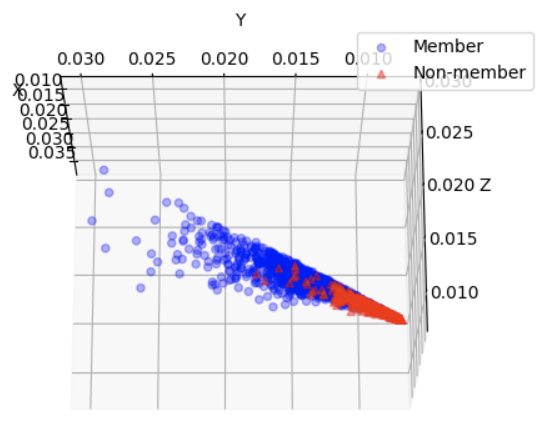}
\label{encoder}
}
\subfigure[avgpool.]{
\includegraphics[width=0.3\linewidth]{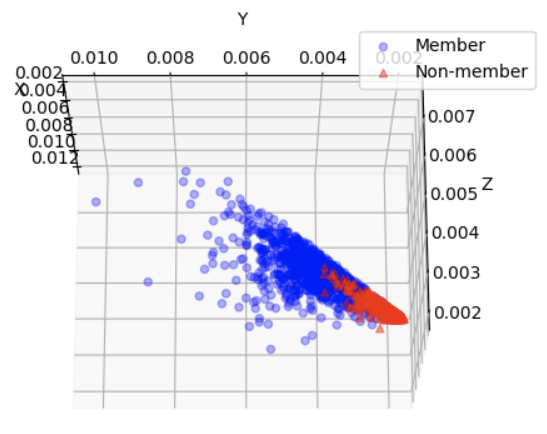}
\label{loss-com2}
}
\subfigure[layer4.]{
\includegraphics[width=0.3\linewidth]{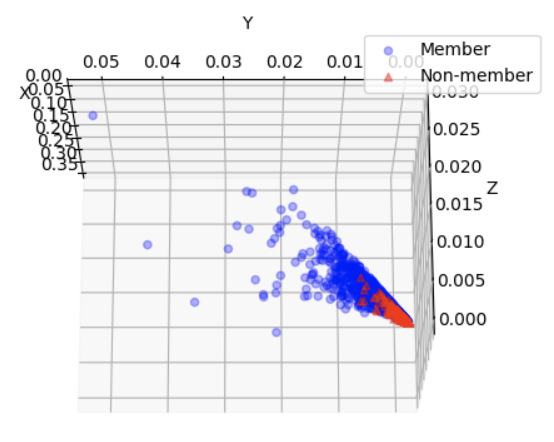}
\label{layer4}
}
\caption{Scatter plot generated using the top 3 predicted probabilities of the output from each layer of the model's encoder.}
\label{fig:san}
\end{figure*}

\begin{table*}[!htb]
	\centering
	\caption{The top 3 confidence scores for member and non-member data across various layers and datasets in CIFAR10.}
	\begin{tabular}{ccc}
		\hline
		& Member & Non-member\\
		\hline
		Encoder & [68.1\%, 14.8\%, 5.9\%] & [18.1\%, 7.7\%, 5.1\%] \\ 
		Avgpool & [1.5\%, 1.0\%, 0.7\%] & [0.3\%, 0.3\%, 0.2\%]  \\ 
		Layer4 & [36.7\%, 11.1\%, 5.4\%] & [4.4\%, 1.2\%, 0.5\%]  \\
		\hline
	\end{tabular}
	\label{tab:confidence}
\end{table*}

\begin{figure*}[!t]
\centering
\subfigure[CIFAR-10.]{
\includegraphics[width=0.3\linewidth]{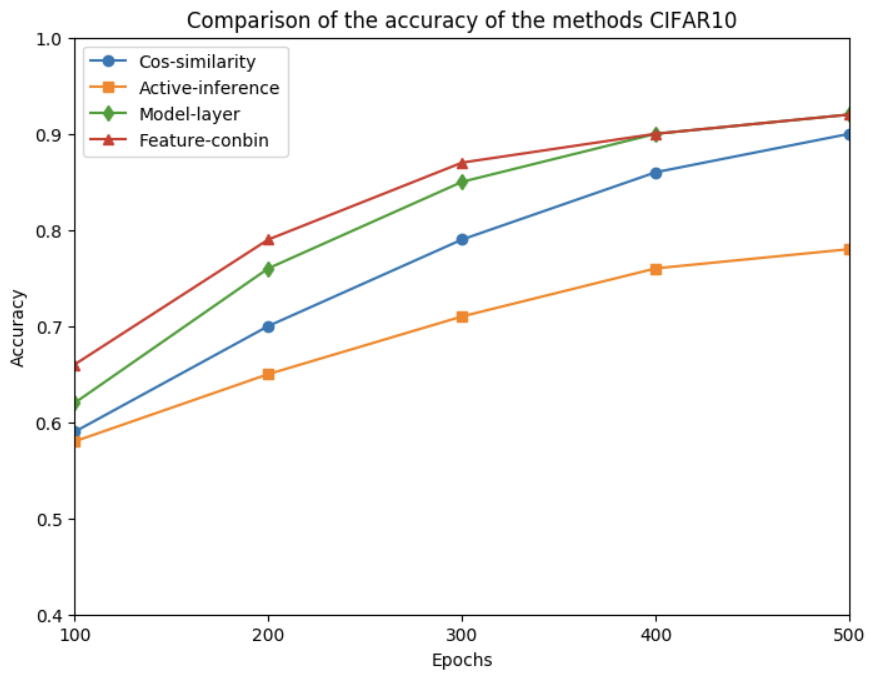}
\label{cifar10_complete}
}
\subfigure[CIFAR-100.]{
\includegraphics[width=0.3\linewidth]{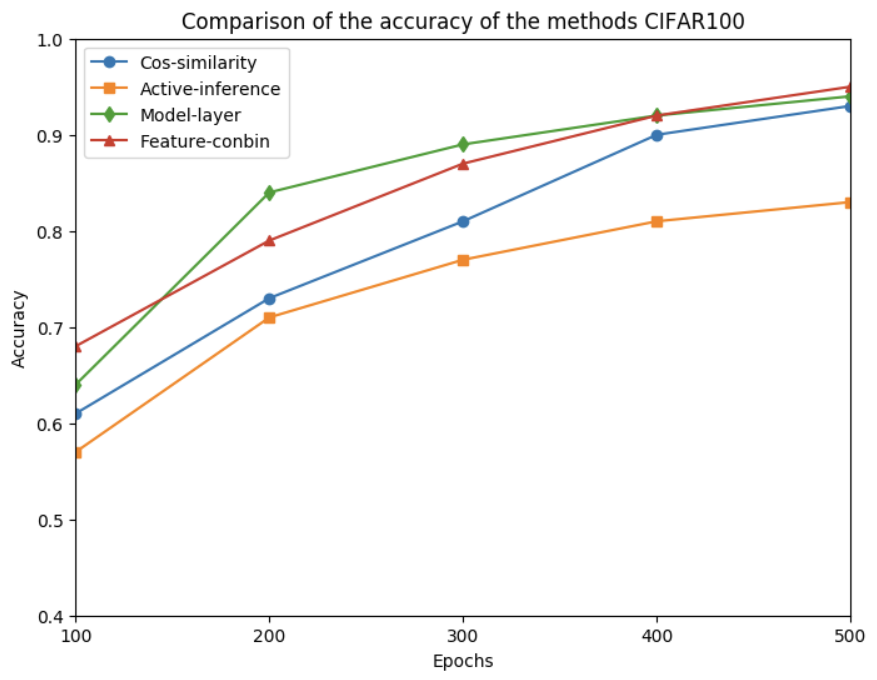}
\label{cifar100_complete2}
}
\subfigure[SVHN.]{
\includegraphics[width=0.3\linewidth]{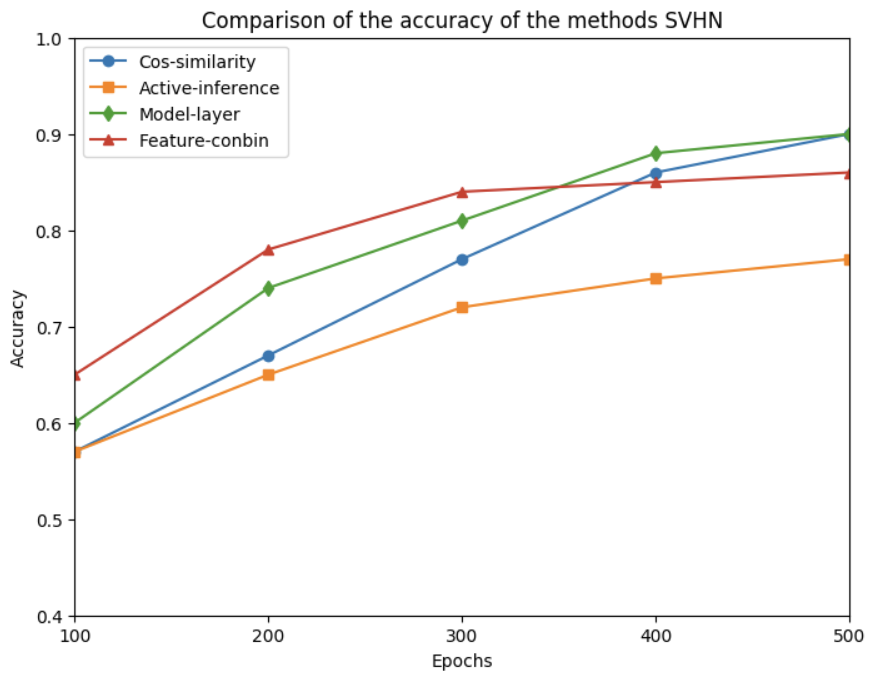}
\label{svhn_complete}
}
\caption{Passive membership inference attack accuracy with feature combinations.}
\label{fig:acc_complete}
\end{figure*}

During our investigation, we analyzed the output of each layer in the encoder for the membership inference attack. We discovered that the output from the final layers of the model can achieve a notably high accuracy when employed as a binary classifier. The success rates of the membership inference attack for each layer of the neural network within the encoder of the extracted model, in relation to the accuracy of the initial passive membership inference attack, are depicted in Figure~\ref{fig:Contrast}.
It reveals that the inference accuracy achieved using this approach surpasses that of the initial method. To further analyze this discrepancy, we conducted a comparison in Table~\ref{tab:confidence}, showcasing the top 3 confidence levels for the mean prediction confidence of each layer within the model. 

Additionally, a 3D distribution plot of the top 3 confidence levels for member and non-member data is presented in Figure~\ref{fig:san}. Notably, the initial layers exhibit minimal variation in overfitting, leading to the selection of layer 4, avgpool, and the encoder's final output for the spatial plot reconstruction.

The visualizations clearly highlight the enhanced capability of this methodology in detecting model overfitting. Therefore, we recommend extracting the internal parameters of the model for an initial risk assessment when evaluating the privacy implications of self-supervised learning models based on contrast learning. This approach can thus facilitate early detection of privacy risks. Furthermore, post 500 rounds of training, this method demonstrates the highest performance among member inference techniques.

\subsubsection{Passive membership inference attack with feature combinations}
In this approach, we integrated the highest prediction probability, data loss, and cosine similarity of the data into a three-dimensional array, significantly enhancing the accuracy of inference. For example, at the 100th epoch depicted in Figure~\ref{fig:acc_complete}, the accuracy rates of alternative methods typically hovered around 60\%. By utilizing cosine similarity, loss, and prediction confidence for linear segmentation, the accuracy rate at the initial 100 epochs surpassed that of the other four methods.

\begin{figure}[!t]
	\centering
	\includegraphics[width=0.3\textwidth]{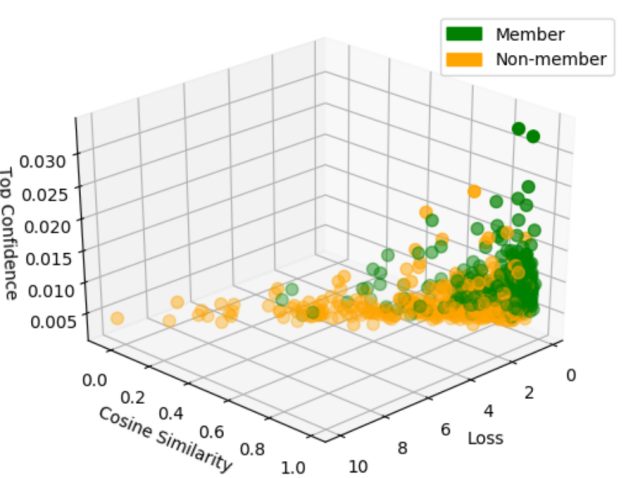}	
	\caption{Scatter plot of the data classified in a three-dimensional space.}
	\label{fig:dis1}
\end{figure}
\begin{table}[!t]
	\centering
	\caption{Classifier Performance Metrics}
	\begin{tabular}{llccc}
		\hline
		Dataset & Classifier & Accuracy & Precision & Recall \\ \hline
		\multirow{3}{*}{cifar10} & SVM & 0.854 & 0.857 & 0.872 \\ 
		& LR & 0.854 & 0.843 & 0.868 \\ 
		& LDA & 0.912 & 0.942 & 0.890 \\ \hline
		\multirow{3}{*}{cifar100} & SVM & 0.912 & 0.859 & 0.100 \\ 
		& LR & 0.902 & 0.846 & 0.100 \\ 
		& LDA & 0.951 & 0.946 & 0.963 \\ \hline
		\multirow{3}{*}{svhn} & SVM & 0.757 & 0.697 & 0.963 \\ 
		& LR & 0.747 & 0.688 & 0.963 \\ 
		& LDA & 0.864 & 0.836 & 0.927 \\ \hline
	\end{tabular}
	\label{tab:classifier_acc}
\end{table}

Figure~\ref{fig:dis1} displayed a three-dimensional scatter plot, enabling a more distinct differentiation between member and non-member data through feature analysis across the three dimensions. Upon comparing the accuracy rates of the aforementioned four methods, it was evident that employing three-dimensional feature data for linear division by the classifier yields a high inference accuracy early in the training process, particularly when overfitting is not pronounced. This underscored the model's privacy risks and offers a refined direction for subsequent defense strategies.

Regarding Membership Inference Attack (MIA), the threat model amalgamated multiple classifiers to identify the optimal performance as the threat model. Table~\ref{tab:classifier_acc} showcased the typical classifier performances such as Support Vector Machine (SVM), Logistic Regression (LR), and Linear Discriminant Analysis (LDA). The classifiers exhibited consistent accuracy levels, suggesting the feasibility of employing a single classifier. Furthermore, plotting the scatter plot of the classifier's data classification using the model-generated features from CIFAR-10 training via FCL revealed that despite the model's moderate overfitting, the classifier trained with these three features maintained exceptional performance, showcasing a notable disparity in the distribution of member and non-member data in space.

\subsubsection{Active membership inference attack}
\begin{figure}[!t]
\centering
\subfigure[Dataset ascent loss.]{
\includegraphics[width=0.46\linewidth]{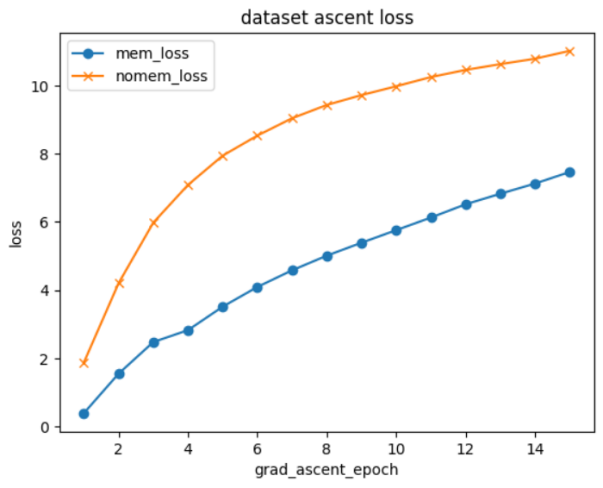}
\label{asc1}
}
\subfigure[Before/after gradient descent.]{
\includegraphics[width=0.46\linewidth]{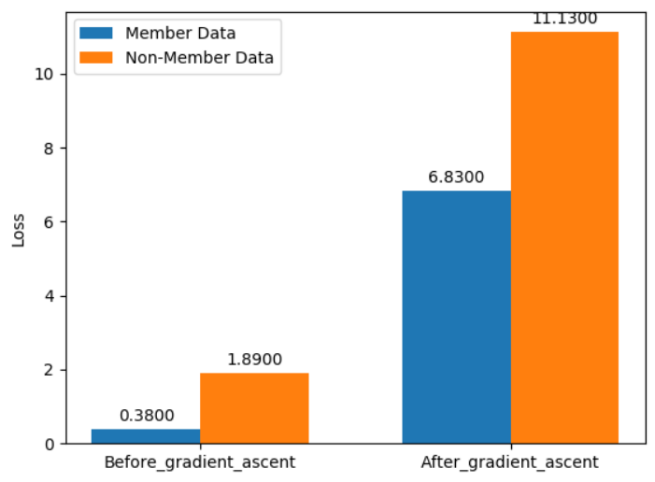}
\label{loss-com2}
}
\caption{Using the model to perform a gradient ascent operation on the data to be inferred, the loss rises with the number of rounds performed.}
\label{fig:ascent_loss}
\end{figure}

\begin{figure}[!t]
\centering
\subfigure[Member data.]{
\includegraphics[width=0.46\linewidth]{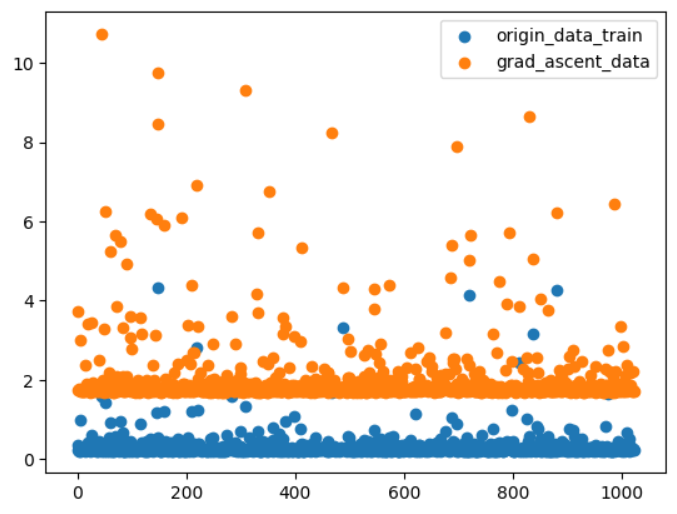}
\label{mem_loss}
}
\subfigure[Non-member data.]{
\includegraphics[width=0.46\linewidth]{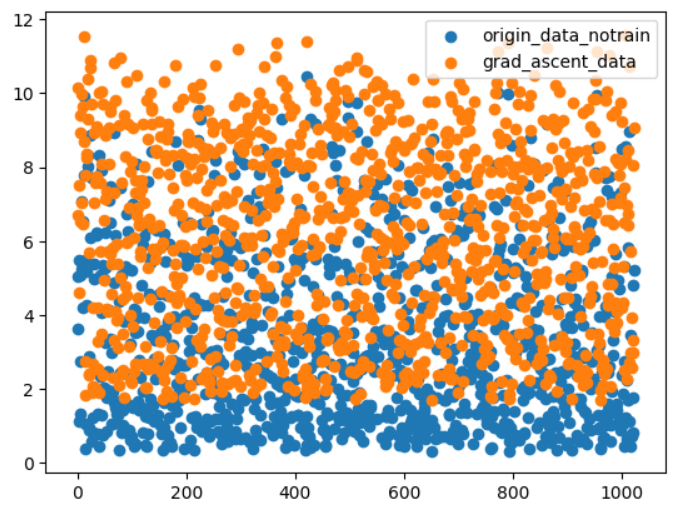}
\label{nonmem_loss}
}
\caption{Loss distributions of member and non-member data.}
\label{fig:data_loss}
\end{figure}

\begin{figure}[!t]
\centering
\subfigure[Accuracy v.s. threshold.]{
\includegraphics[width=0.46\linewidth]{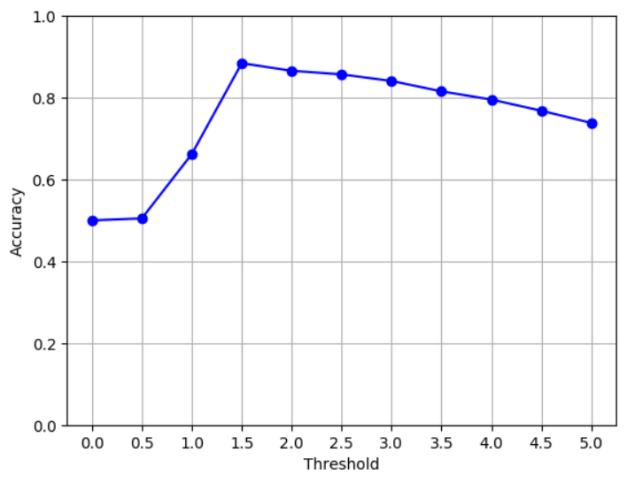}
\label{threshold}
}
\subfigure[Accuracy v.s. training epochs.]{
\includegraphics[width=0.46\linewidth]{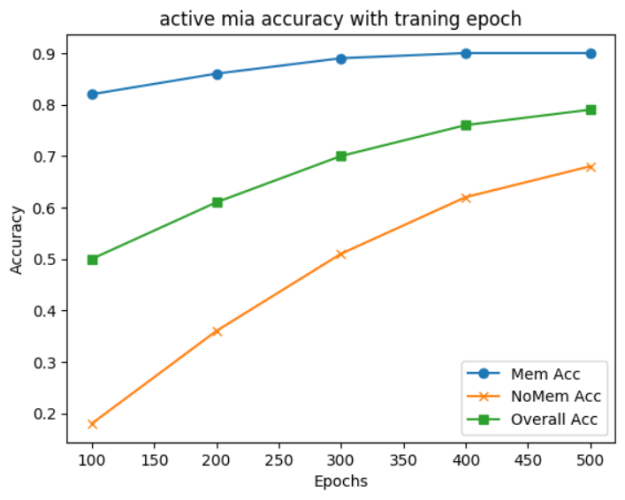}
\label{active_acc}
}
\caption{Active membership inference attack accuracy with respect to different thresholds and training epochs.}
\label{fig:dynamic_loss}
\end{figure}

In the active membership inference attack, we initially explored a static approach. This involved conducting a gradient ascent operation on a batch of data for inference and observing the resulting increase in data loss. Figure~\ref{fig:ascent_loss} illustrates that the loss of non-member data significantly surpasses that of member data.

Figure~\ref{fig:data_loss} depicts the loss trends of member and non-member data pre and post gradient ascent. Figure~\ref{fig:data_loss}(a) showcases the data loss before and after gradient ascent for member data, while Figure~\ref{fig:data_loss}(b) displays the same for non-member data. It is evident that member data exhibits a more concentrated distribution post-gradient ascent, whereas non-member data appears more chaotic with varying data point losses, albeit showing an overall increase in data loss post-ascent.

Subsequently, a threshold value can be established for the loss increase, classifying data exceeding this threshold as non-member data and vice versa. By utilizing this threshold, data can be segmented into member and non-member categories. In Figure~\ref{fig:dynamic_loss}, accuracy is computed separately for each category, with the final accuracy being the average of both. With an optimal threshold selection, the success rate of the active method can exceed 80\%. However, adjustments to the epoch and learning rate of the gradient ascent operation are necessary. Models with higher overfitting require higher learning rates and epochs for effective gradient ascent, while models with less overfitting benefit from lower rates and epochs.

\begin{figure}[!t]
\centering
\subfigure[Model loss with the gradient ascent method.]{
\includegraphics[width=0.46\linewidth]{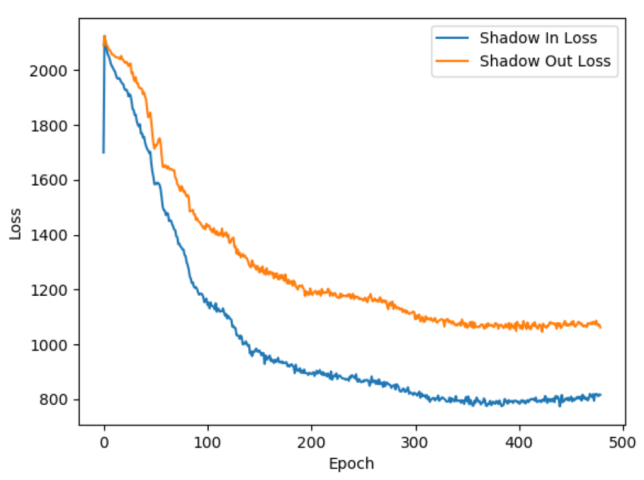}
\label{active_loss_compare}
}
\subfigure[Model losses across different aggregation methods.]{
\includegraphics[width=0.46\linewidth]{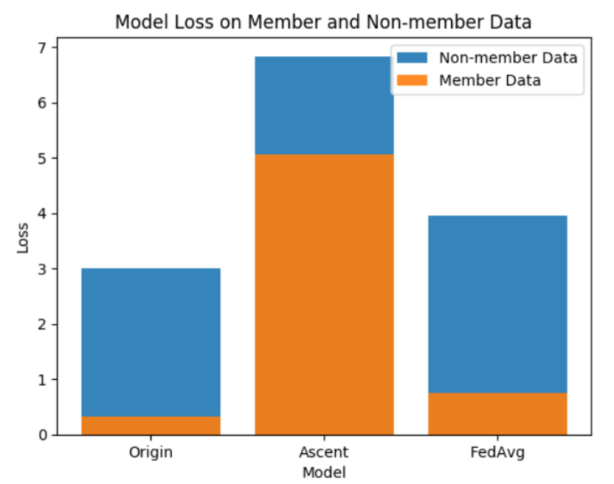}
\label{loss_memandnonmem}
}
\caption{
The loss incurred by the active membership inference attack on FCL, which involves executing a gradient ascent operation on the local model in each round and then utilizing the aggregated model to assess the loss on the inferred data.}
\label{fig:dynamic_active}
\end{figure}

Integrating the gradient ascent operation into federated learning involves performing this operation on the aggregated model per round. This ensures consistency in optimizer settings and learning rates between training and aggregation. Theoretical implications suggest that membership data will notably reduce the model's loss post-aggregation, while non-membership data will experience a slower reduction, potentially increasing if the model becomes significantly overfitted. Figure~\ref{fig:dynamic_active} demonstrates the loss changes from gradient ascent to model aggregation, highlighting the importance of matching the epochs and learning rates during retraining to offset the effects of aggregation.

In essence, the FCL process entails conducting gradient ascent operations on local models each round, aggregating these models, and assessing the loss of inferred data. This method is ideal for discerning membership status within a dataset, as the loss curve of the data directly indicates whether it belongs to the membership category. By observing the loss curve evolution during training, one can easily differentiate between member and non-member data without the need for direct comparison, making the inference results straightforward.

\section{Conclusion and Future Work}
\label{sec:conclusion} 
This paper introduces a pioneering investigation into passive and active membership inference attacks on federated contrastive learning. Our research substantiates the potential data leakage vulnerability inherent in federated contrastive learning. Experimental findings demonstrate that an excessively overfitted model resulting from improper parameter choices or highly skewed data distribution among clients poses heightened privacy risks. The privacy concerns within federated contrastive learning warrant significant attention. In the future research, incorporating defensive measures like differential privacy and early stopping enhances the generalization capability of the federated contrastive learning model, fortifying it against membership inference attacks and privacy breaches. 

\section{Acknowledgments}
This work is supported by National Natural Science Foundation of China (No. 61802383), Research Project of Pazhou Lab for Excellent Young Scholars (No. PZL2021KF0024), and Guangzhou Basic and Applied Basic Research Foundation (No. 202201010330, No. 202201020162).


\begin{IEEEbiographynophoto}{Kongyang Chen}
received the PhD degree in computer science from the University of Chinese Academy of Sciences, China. He is currently an Associate Professor at Guangzhou University, China. Before that, he was a Postdoc Fellow at the Hong Kong Polytechnic University, Hong Kong, China. From 2014 to 2018, he was an Assistant Professor at Shenzhen Institutes of Advanced Technology, Chinese Academy of Sciences, China. His research interests are artificial intelligence, edge computing, blockchain, IoT, etc.
\end{IEEEbiographynophoto}

\begin{IEEEbiographynophoto}{Wenfeng Wang}
is a master student at Guangzhou University. His research interests are federated learning and machine unlearning.
\end{IEEEbiographynophoto}

\begin{IEEEbiographynophoto}{Zixin Wang}
is a master student at Guangzhou University. His research interests are federated learning and machine unlearning.
\end{IEEEbiographynophoto}

\begin{IEEEbiographynophoto}{Zhipeng~Li}
is an engineer at Jiangxi Qiushi Academy for Advanced Studies, China. He received his master degree at Institute of Artificial Intelligence, Guangzhou University, China. His main research interests are deep learning and security \& privacy.
\end{IEEEbiographynophoto}
  
\begin{IEEEbiographynophoto}{Yao~Huang}
is with Guangzhou College of Commerce, Guangzhou, China. He received his master degree at Institute of Artificial Intelligence, Guangzhou University, China. His main research interests are deep learning and security \& privacy.
\end{IEEEbiographynophoto}

\end{document}